\documentclass[sigconf, nonacm]{acmart}

\usepackage{algorithm}
\usepackage{algpseudocode}
\usepackage{tabularx}
\usepackage{colortbl}
\usepackage{multirow}
\usepackage{threeparttable}
\usepackage{enumitem}
\usepackage{microtype}
\usepackage{placeins}
\usepackage{flafter}
\usepackage{balance}
\usepackage{siunitx}
\usepackage{xspace}

\newcolumntype{L}[1]{>{\raggedright\arraybackslash}p{#1}}

\newcommand\vldbdoi{XX.XX/XXX.XX}

\newcommand\vldbavailabilityurl{https://github.com/AgentCombo/Mandol}
\newcommand\vldbpagestyle{plain}

\begin{document}

\title{Mandol: An Agglomerative Agent Memory System for Long-Term Conversations}

\author{Yuhan Zhang$^{\dagger}$, \ Zhiyuan Guo$^{\dagger}$, \ Ziheng Zeng$^{\dagger}$, \ Wei Wang$^{\dagger}$, \ Wentao Wu$^{\ddagger}$, \ Lijie Xu$^{\dagger}$}
\affiliation{
\institution{$^{\dagger}$Institute of Software, Chinese Academy of Sciences \quad $^{\ddagger}$Microsoft Research \\ xulijie@iscas.ac.cn} 
\city{\texttt{\textcolor{blue}{Github: https://github.com/AgentCombo/Mandol}}}
\country{}
}

\begin{abstract}
Long-term conversational agents need to remember and query cross-session, multi-typed information with complex correlations. Existing agent memory systems rely on heterogeneous vector and graph databases, which fragment memory information and cause high cross-database I/O latency. For retrieval, common RAG-style methods tend to introduce noise, miss correlated clues, and lack token budget control, degrading LLM accuracy and efficiency.

We propose Mandol, an agglomerative memory system that consolidates fragmented memory representations and storage into a unified memory-native architecture. Its core components include: (1) a \emph{hierarchical memory model} that organizes memory into a basic layer representing raw memory information and a high-level abstract layer that agglomerates basic memories into traceable abstract memories, both uniformly represented as structured semantic graphs; (2) an \emph{agglomerative semantic data structure} combining \emph{SemanticMap} and \emph{SemanticGraph}, which natively fuses key-value, vector, and graph structures and provides unified hybrid retrieval operators to eliminate cross-database I/O; and (3) a \emph{quantitative query mechanism} with query-adaptive routing, quantitative denoising and conflict resolution, and token-constrained context generation, all without involving LLMs during retrieval. Experiments on two widely used long-term conversation benchmarks, LoCoMo and LongMemEval, show that Mandol achieves the best overall accuracy among representative agent memory systems. For performance comparison, Mandol also obtains a \(5.4\times\) retrieval speedup and a \(4.8\times\) insertion speedup under 10 QPS concurrent load, while still maintaining low latency on consumer-grade hardware.
\end{abstract}

\maketitle

\pagestyle{\vldbpagestyle}


\section{Introduction}
\label{sec:intro}

LLM conversational agents~\cite{LLM-agent-survey1, memory-survey1, AI-agent-memory-survey} are shifting from single-turn Q\&A to continuous, long-term interactions in applications such as customer service~\cite{yisha2025customersupport}, personal assistance~\cite{Personal-assistance}, and medical diagnosis~\cite{li2024agenthospital, li2023chatdoctor}.
Agent memory modules need to store and query cross-session, highly correlated information such as dialogues, user intents, events, and entity states with low latency.
For instance, a customer service agent tracks dialogue history, orders, product information, and user sentiment, while a medical assistant agent connects symptoms, historical records, and test results.

Memory queries in long-term conversational agents are complex and diverse, ranging from simple fact extraction to reasoning-intensive queries that keyword or similarity matching cannot handle.
These complex queries often involve temporal relationships, multi-hop reasoning, or evolving user states, and they become especially vulnerable to noise or missing evidence when memories are fragmented or span long periods.
(1) \textit{Temporal retrieval} heavily depends on event order and time spans. A query such as ``which hotels did I book during my trip to Northern Europe last summer, and what was the total cost?'' requires precise tracing and aggregation across sessions.
(2) \textit{Cross-session multi-hop reasoning} demands logical inference over fragmented sessions. If a user mentions a seafood allergy in one session and later reports a skin rash, the system must connect the allergy to the symptom.
(3) \textit{Memory state update and denoising} must track changing user preferences over time. When a housing preference shifts from ``suburban apartments'' to ``apartments in urban school districts'', the system must generate recommendations based solely on the latest constraints.
Long-context studies further show that simply placing more history into the prompt is unreliable: models can miss evidence in the middle of long contexts or degrade when irrelevant tokens accumulate~\cite{liu2023lostmiddle,levy2024sametask}.

These demands introduce three core challenges in current agent memory systems like Mem0~\cite{chhikara2025mem0}, Zep~\cite{rasmussen2025zep}, and MemOS~\cite{memos2025}:
\begin{itemize}[leftmargin=*]
    \item \emph{Unified memory representation.} Long-term conversational memory has diverse types, complex relationships, and dynamic evolution patterns. Vector embeddings cannot explicitly express logical structures or temporal relationships, while knowledge graphs rely on rigid schemas that struggle to capture semantic similarity and flexibly represent heterogeneous memory information.
    \item \emph{Efficient storage and hybrid querying.} Without a unified representation model, current systems combine vector databases, graph databases, and other stores, forcing hybrid queries to involve cross-database orchestration with high I/O and serialization overhead, which fails to meet low-latency interaction demands.
    \item \emph{Accurate retrieval with token budget constraints.} Common RAG-style retrieval relies on similarity matching, which tends to introduce noise, miss associative clues, and lack token-budget control. As a result,
the retrieved memories can be incomplete, conflicting, or redundant, causing simple queries to waste context on noisy memories and complex multi-hop queries to lose critical evidence due to context-window truncation.
Prompt compression and streaming attention mechanisms alleviate some of this pressure~\cite{jiang2024longllmlingua,xiao2024streamingllm,munkhdalai2024infiniattention}, but they do not determine which long-term memories are fresh, non-conflicting, and evidence-grounded for a given query.
\end{itemize}

To systematically address these challenges, 
we propose Mandol, an agglomerative memory system that consolidates fragmented memory representations and storage into a unified memory-native architecture.
Its core modules and techniques include:

(1) \textit{Hierarchical memory model.}
We organize memory into a basic memory layer and a high-level abstract memory layer, both uniformly represented as structured semantic graphs.
The basic layer represents raw memory through memory units, spaces, and explicit/implicit relationships.
The abstract layer automatically agglomerates basic memories into compact abstractions, including episodic memory (event chains), semantic memory (entity graphs), and emotional memory (user preferences). These abstractions maintain traceable links back to the original basic memories, ensuring evidence grounding while supporting abstract reasoning.

(2) \textit{Agglomerative semantic data structures.}
We design a unified in-memory data structure combining \emph{SemanticMap} and \emph{SemanticGraph}, which natively fuses key-value, vector, and graph structures and provides unified hybrid retrieval operators to eliminate cross-database I/O. This agglomerative design removes the I/O latency of heterogeneous storage.
The data structures also connect to an underlying persistent database for cold or long-term memory.

(3) \textit{Quantitative retrieval mechanism.}
We replace the traditional RAG-style recall-then-rank paradigm with a pipeline of query-adaptive routing, quantitative denoising and conflict resolution, and token-constrained context generation.
Query-adaptive routing dynamically selects and queries relevant memory sources based on query intent.
Quantitative denoising and conflict resolution then remove noise and contradictory information from different sources.
Finally, a compact high-quality context is generated under token constraints by jointly optimizing relevance and diversity, all without invoking LLMs during retrieval.

Compared to representative open-source memory systems, including Mem0, Zep, MemU~\cite{nevamind2026memu}, MemOS, and EverMemOS~\cite{evermemos2026}, Mandol achieves the highest overall accuracy on LoCoMo (92.21\%) and LongMemEval (88.40\%).
It also obtains the best system performance: under a 10\,QPS concurrent serving setting, Mandol reduces Search mean latency by $5.4\times$ and Add mean latency by $4.8\times$ over the fastest baselines, and it maintains lower latency than existing systems even on consumer-grade hardware.
Mandol is available at \url{https://github.com/AgentCombo/Mandol}.

In summary, this paper makes the following contributions.
\begin{itemize}[leftmargin=*]
    \item We propose a hierarchical memory model that uniformly represents basic and abstract memories through structured semantic graphs, and agglomerates basic memories into traceable high-level abstractions, solving the unified representation problem.
    \item We design an agglomerative semantic data structure combining \emph{SemanticMap} and \emph{SemanticGraph}, which fuses key-value, vector, and graph structures with atomic hybrid retrieval operators, eliminating memory fragmentation and high-latency queries.
    \item We propose a quantitative retrieval mechanism that integrates routing, denoising, conflict resolution, and token-constrained context generation to achieve accurate and token-efficient retrieval without LLM intervention.
    \item Mandol achieves the state-of-the-art accuracy on LoCoMo and LongMemEval, while delivering the lowest latency in both server and consumer settings, verifying its effectiveness, efficiency, and stability in long-conversation scenarios.
\end{itemize}
\section{Related Work}
\label{sec:related-work}

Long-term conversational memory has become a fundamental requirement across diverse agents. Recommendation agents such as InteRecAgent~\cite{huang2023interecagent} track users' intents and preferences across sessions to generate accurate recommendations. Social interaction agents like Generative Agents~\cite{park2023generative} and RoleLLM~\cite{RoleLLM} store dialogue histories, behaviors, events, or user profiles to preserve character consistency. Medical expert agents such as AgentHospital~\cite{li2024agenthospital} and ChatDoctor~\cite{li2023chatdoctor} retain extensive doctor-patient dialogues, records, and domain knowledge to support multi-hop reasoning and fact tracing. All these scenarios share a common demand: the memory system must efficiently store and accurately retrieve long-term, cross-session, and multi-typed information with low query latency.

\subsection{Memory Representation and Storage}
\label{subsec:rw-systems}

Current agent memory systems face significant challenges in both representation and storage. As summarized in Table~\ref{tab:system_comparison}, Mem0~\cite{chhikara2025mem0} primarily relies on vector embeddings for similarity matching and can optionally use a knowledge graph to capture entity relationships. Zep~\cite{rasmussen2025zep} builds a temporal knowledge graph that tracks timestamps for facts and relationships as they evolve over time. MemOS~\cite{memos2025} and EverMemOS~\cite{evermemos2026} represent memory using text vectors along with tree-structured or scenario-level summaries. However, these designs remain fragmented across vector and graph modalities, lacking a unified memory model capable of covering diverse memory types along with both structural and semantic dimensions.
For storage, these systems commonly adopt a heterogeneous composition of vector databases~\cite{Milvus,qdrant}, graph databases~\cite{angles2008survey,neo4j}, and other specialized stores.
This fragmented architecture forces hybrid queries to cross database boundaries and requires application-level orchestration, incurring high serialization I/O overhead that hinders low-latency real-time interaction.

\begin{table*}[t]
\centering
\caption{System-level comparison of agent memory systems.}
\label{tab:system_comparison}
\footnotesize
\setlength{\tabcolsep}{3pt}
\begin{tabularx}{\textwidth}{@{}L{0.10\textwidth}L{0.23\textwidth}L{0.23\textwidth}>{\raggedright\arraybackslash}XL{0.14\textwidth}@{}}
\toprule
\textbf{System} &
\textbf{Memory Organization} &
\textbf{Storage} &
\textbf{Retrieval} &
\textbf{I/O overhead} \\
\midrule
    Mem0~\cite{chhikara2025mem0} &
    Text vectors + optional knowledge graph &
    VectorDB + optional GraphDB &
    Vector search + optional graph search &
    Medium \\

    Zep~\cite{rasmussen2025zep} &
    Text vectors + temporal knowledge graph &
    GraphDB + Lucene &
    Graph search + rerank &
    High \\

    MemOS~\cite{memos2025} &
    Text vectors + memory summaries &
    VectorDB + GraphDB &
    Vector search + graph-based lookup &
    High \\

    EverMemOS~\cite{evermemos2026} &
    Text vectors + memory summaries &
    VectorDB + RDBMS + Markdown &
    Vector search + multi-round retrieval &
    High \\ \hline

    Mandol (ours) &
    Basic + high-level memories, represented as structured semantic graphs &
    In-memory SemanticMap/Graph + In-process DuckDB &
    Token-constrained quantitative retrieval &
    Low \\ 
    \bottomrule
\end{tabularx}
\end{table*}

Meanwhile, the database community has also investigated unified data management across vector and structured data. Several recent systems integrate vector search into relational databases~\cite{SingleStore-V, GaussDB-V, PostgreSQL-V}, and extend graph-based approximate nearest neighbor (ANN) indexes~\cite{acorn2024, graph-based-search-eval}. These efforts primarily focus on optimizing vector search performance and integrating it with relational query processing. However, complex conversational memory contains multi-typed information (e.g., dialogues, events, user preferences), intricate correlation structures (e.g., temporal order, causality, entity associations), and abstractions (e.g., episodic and semantic memories), which cannot be easily stored or queried by existing solutions.

Beyond explicit storage, recent works also explore implicit
approaches~\cite{AI-agent-memory-survey}. Parametric memory
approaches~\cite{Retroformer, behrouz2024titans} internalize knowledge and experience
into model parameters (weights). Latent memory approaches~\cite{yang2024memory3, MemGen} leverage internal model representations like dynamic hidden states and KV caches during inference. However, parametric memory suffers from limited interpretability and high update costs, while latent memory typically requires costly pretraining or multi-stage training to construct memory representations. Explicit storage thus remains necessary in practical systems, and our work focuses on this direction.

\subsection{Memory Retrieval Methods}
\label{subsec:rw-retrieval}

Most current agent memory systems adopt RAG-style retrieval, which typically retrieves content through vector similarity~\cite{lewis2020retrieval,gao2024ragsurvey,gao2024modularrag}.
GraphRAG~\cite{darren2024graphrag}, RAPTOR~\cite{sarthi2024raptor}, and LightRAG~\cite{guo2024lightrag} build graph or tree structures over retrieved evidence to improve global or multi-hop reasoning. HippoRAG~\cite{gutierrez2024hipporag} builds a knowledge graph index with personalized PageRank for multi-hop retrieval. A-MEM~\cite{xu2025amem} constructs a Zettelkasten-based knowledge network supporting memory evolution. Other works focus on system-level retrieval optimization, such as skyline-based chunk retrieval with query decomposition for long-context QA~\cite{LLM-QA} and workload-aware vector partitioning for high-throughput hybrid queries over knowledge graphs~\cite{RAG-KG}.

For long-term conversations, however, these methods suffer from two key limitations. First, similarity-based retrieval easily introduces noise, conflicts, and redundancy, especially when memory states evolve and contradict across sources. Second, token consumption remains uncontrolled: the system cannot adjust token usage based on query complexity, so simple queries may retrieve excessive results and waste tokens, while complex multi-hop queries risk truncation and lose critical evidence chains.

In summary, how to achieve unified memory representation, efficient storage with low-latency hybrid query support, and accurate token-efficient retrieval remains an open challenge for long-term conversational agents.

\section{Mandol System Design}
\label{sec:system-design}

We design Mandol to address three key questions: (1) How can we uniformly represent long-term conversational memory that is multi-typed, correlated, and dynamically evolving? (2) How can we store such memory in a unified form and support low-latency hybrid queries? (3) How can we retrieve memories accurately and efficiently under a limited token budget? We propose a hierarchical memory model (Section~\ref{subsec:hierarchical-memory-model}), a unified storage system based on agglomerative semantic data structures (Section~\ref{subsec:semantic-data-structures}), and a quantitative query mechanism (Section~\ref{subsec:smart-quantitative-querying}).

\begin{figure*}[t]
    \centering
    \includegraphics[width=0.9\textwidth]{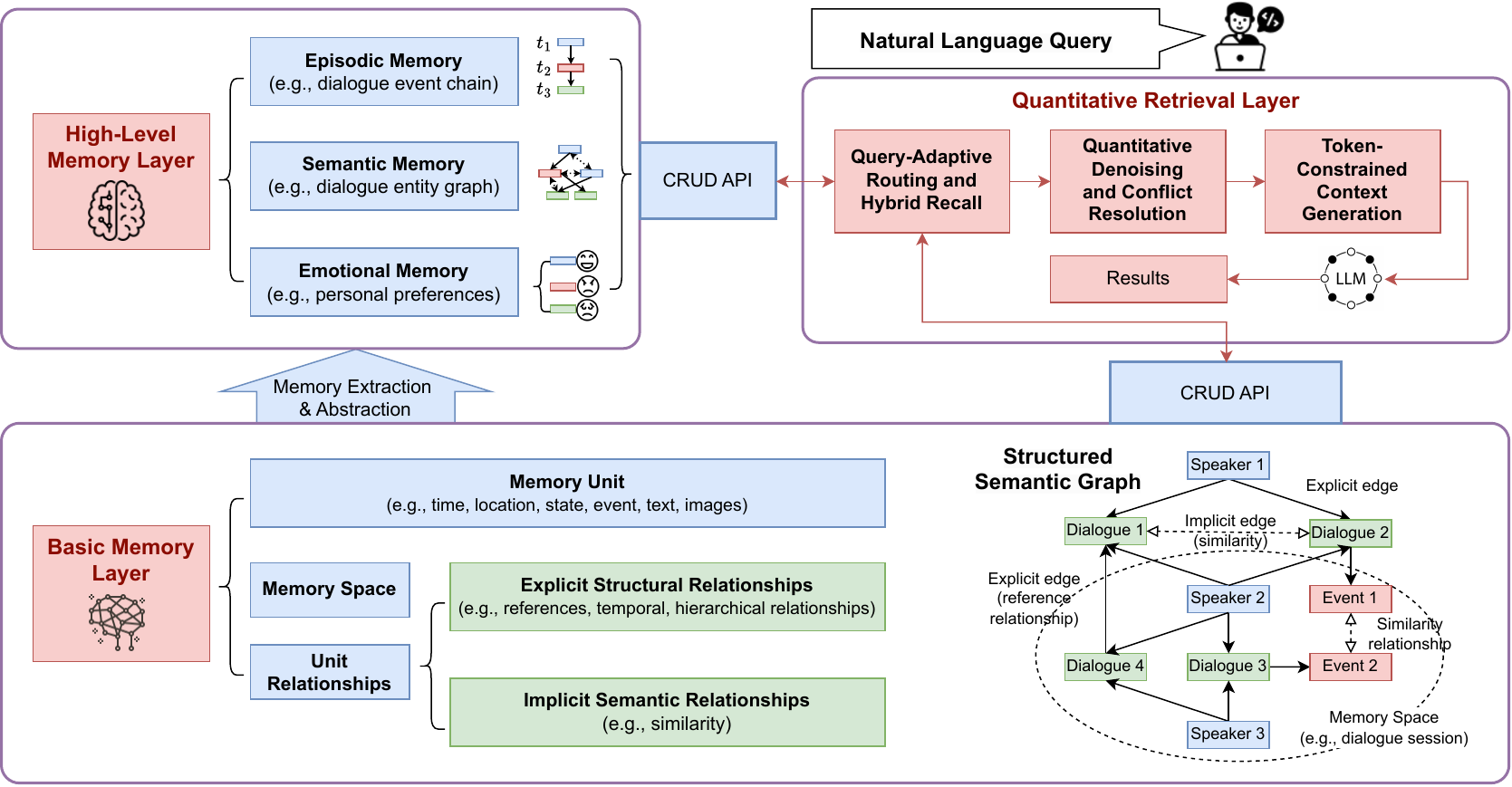}
    \Description{Overview of Mandol's hierarchical memory model.}
    \caption{Overview of Mandol's hierarchical memory model.}
    \label{fig:memory_model}
\end{figure*}

\subsection{Hierarchical Memory Model}
\label{subsec:hierarchical-memory-model}

To address the unified representation problem, Mandol builds a hierarchical memory model with three layers: a \textit{basic memory layer}, a \textit{high-level abstract memory layer}, and a \textit{quantitative query layer}. The core idea is that both the basic and abstract layers use a structured semantic graph for unified representation, differing only in organization and abstraction level. The basic layer directly organizes raw interaction information; the abstract layer leverages an LLM to automatically extract and agglomerate traceable abstract memories; the query layer performs query-adaptive routing and quantitative retrieval over both layers (detailed in Section~\ref{subsec:smart-quantitative-querying}). Figure~\ref{fig:memory_model} illustrates this layered model, in which raw conversational evidence is transformed into traceable abstract memories and then accessed by the query layer.

\subsubsection{Basic Memory Layer}
This layer uniformly represents raw interaction information. It constructs a structured semantic graph by decomposing raw information into three core elements: memory units, memory spaces, and explicit/implicit memory relationships.

A \textit{memory unit} is a fine-grained unit that encapsulates raw information such as dialogue text, user intent, and temporal or spatial metadata. A pretrained model encodes the raw information into a unified semantic vector, which is stored alongside the raw information and metadata within the unit, enabling uniform storage and semantic alignment of multi-typed memory information.

A \textit{memory space} provides logical isolation at multiple granularities: it can group memory units by session, task, or topic, and also organize abstract memories (e.g., episodic, semantic, emotional) into distinct spaces. A single unit can belong to multiple spaces through metadata-based space membership, allowing controlled cross-session retrieval without merging unrelated contexts.

Memory relationships consist of \textit{explicit structural edges} (e.g., temporal order, entity reference, state update) and \textit{implicit semantic edges} (semantic similarity). Explicit edges are added directly through rule parsing to guarantee precise structural relationships. Implicit edges are computed on demand during queries via semantic indexes, avoiding redundant edge storage while preserving the flexibility of semantic retrieval. Together, these elements form a structured semantic graph that supports incremental updates and provides a unified data view with both structural precision and semantic flexibility.

\subsubsection{High-level Abstract Memory Layer} This layer leverages an LLM to agglomerate compact abstract memories from the basic memory layer through an abstraction-linking procedure. The LLM first extracts key information and organizes it into episodic memory as event chains, semantic memory as entity graphs, and emotional memory as user preferences. These abstract memories are uniformly represented via the same structured semantic graph: event chains use temporal and causal edges to connect event nodes, entity graphs use reference and attribute edges, and user preferences use state-update edges to track change trajectories. Each abstract node stores \texttt{source\_unit\_uids}, creating traceable links back to the original basic memory units, which allows the system to verify abstract evidence.

To address the fragmentation of raw memory and the loss of context in long conversations, Mandol applies a session-level context enhancement technique during the abstraction process. It augments short dialogue snippets with session-level metadata such as timestamps and topic descriptors, turning isolated utterances into context-rich retrieval units. For instance, an isolated utterance like ``\textit{booked a hutong homestay}'' is enriched into ``\textit{[Time: 2026-05-25 | Topic: Mike's trip to Beijing] Booked a hutong homestay}'', which can boost cross-session retrieval. Mandol then links this event to an earlier session where Mike planned to book the ``\textit{Wangfujing Hotel}'' and abstracts a state-update edge capturing the accommodation evolution from ``\textit{Wangfujing Hotel}'' to ``\textit{hutong homestay}''.

\subsection{Agglomerative Semantic Data Structure}
\label{subsec:semantic-data-structures}

\begin{figure*}[t]
    \centering
    \includegraphics[width=\textwidth]{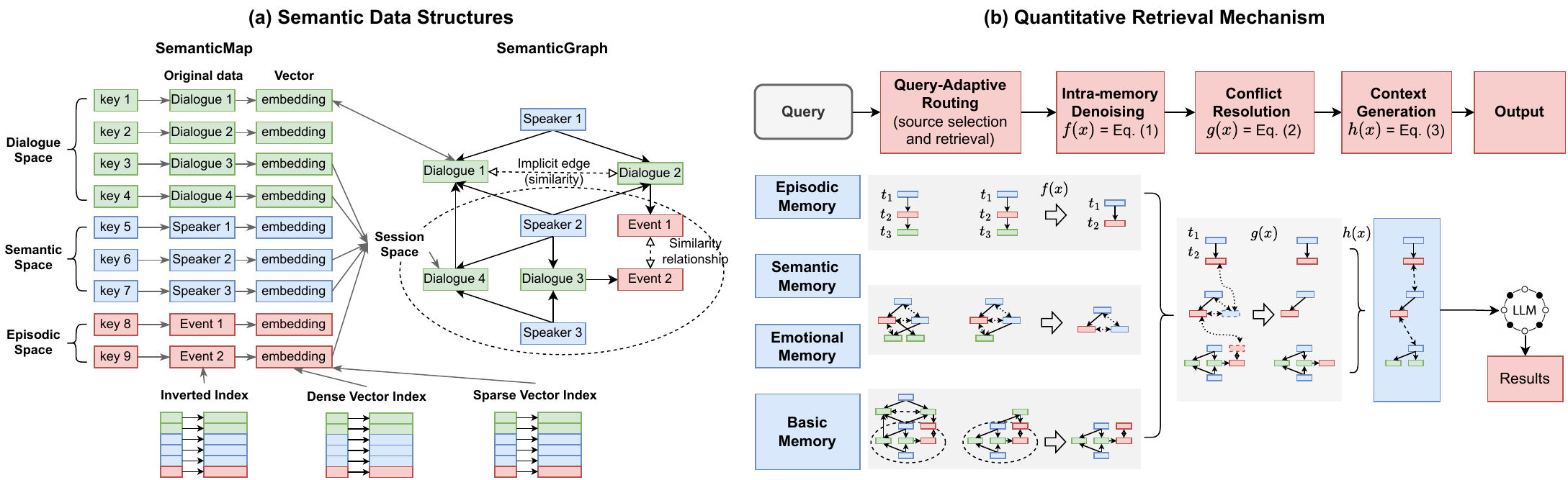}
    \Description{Semantic data structures and smart quantitative retrieval mechanism in Mandol.}
    \caption{The semantic data structures and quantitative retrieval mechanism in Mandol.}
    \label{fig:data_structure_retrieval}
\end{figure*}

Existing systems combine vector databases, graph databases, and other databases, which fragment memory and force hybrid queries to cross database boundaries with high I/O and serialization overhead. Guided by the hierarchical memory model, Mandol replaces this architecture with an agglomerative semantic data structure composed of \textit{SemanticMap} and \textit{SemanticGraph}, which natively fuses key-value storage, vector indexes, and graph topology within a single address space.
Figure~\ref{fig:data_structure_retrieval} shows how these semantic data structures support both unified storage and the downstream quantitative retrieval pipeline. The left side corresponds to the in-memory organization, while the right side corresponds to the retrieval stages introduced in Section~\ref{subsec:smart-quantitative-querying}.

\textit{SemanticMap} combines key-value storage and vector structures to support heterogeneous data storage and semantic queries over memory units. It uses the memory unit ID as the key and records space membership via metadata for memory space isolation. The value stores raw information and the semantic vector. Mandol provides three types of indexes: an inverted index on raw memory information (e.g., dialogue text) for keyword retrieval, and optimized dense and sparse vector indexes for low-latency semantic retrieval.

\textit{SemanticGraph} uses a lightweight adjacency list to store explicit structural edges and obtains implicit semantic relationships on demand from SemanticMap's indexes. SemanticMap and SemanticGraph share the same Memory Unit IDs, so every operation (lookup, filtering, vector search, graph traversal) returns the same identifiers. This makes hybrid retrieval atomic: a memory unit retrieved by dense search can be expanded through graph traversal and filtered by space metadata within the unified data structures. Together, SemanticMap and SemanticGraph form a unified storage view, physically realizing the structured semantic graph and eliminating multi-database fragmentation.

To support flexible retrieval, the system provides a set of atomic hybrid operators that natively operate on SemanticMap and SemanticGraph, covering unit, space, relationship, and multi-hop queries. These operators can be composed into complex retrieval plans and are detailed in Section~\ref{sec:implementation}. The active memory layer connects to an underlying embedded persistence backend, DuckDB, for cold or long-term storage through the asynchronous paging mechanism described in Section~\ref{sec:implementation}.

\subsection{Quantitative Query Mechanism}
\label{subsec:smart-quantitative-querying}

In long-term conversations, RAG-style retrieval based on keyword or semantic matching easily introduces noise, misses multi-hop evidence, and lacks token budget control. Mandol instead models retrieval as high-quality context construction under a finite token budget and addresses these problems through three stages: query-adaptive smart routing, quantitative denoising and conflict resolution, and token-constrained context generation. The overall retrieval flow is shown on the right side of Figure~\ref{fig:data_structure_retrieval}.

\subsubsection{Query-Adaptive Routing}
\label{subsubsec:query-adaptive-routing}

Mandol maintains multiple memory sources (basic, episodic, semantic, and emotional memory) constructed by the hierarchical model. Retrieving from all sources indiscriminately incurs high computational cost and introduces abundant query-irrelevant noise. We adopt a strategy that selects the most relevant memory sources on demand based on query-level semantic features, thereby filtering noise and controlling the candidate budget at the source.

A lightweight intent classifier selects relevant memory sources and allocates per-source candidate budgets using query features such as temporal expressions and entity mentions. For example, time-sensitive queries activate episodic and basic memories while skipping the semantic memory when entity relations are not needed.

Within each activated source, hybrid recall runs BM25 keyword retrieval~\cite{robertson2009probabilistic}, SPLADE sparse expansion~\cite{formal2021splade}, and dense vector search~\cite{johnson2019billion} in parallel. Candidate lists are fused via reciprocal rank fusion~\cite{cormack2009reciprocal}. To recover evidence for multi-hop reasoning, Mandol performs selective subgraph expansion from top-ranked units over SemanticGraph, pruning by edge type and budget. Finally, a lightweight cross-encoder reranks the expanded candidates, producing a semantic relevance score \(S_{ce}(c_i)\) for each candidate \(c_i\) for the next stages.

\subsubsection{Quantitative Denoising and Conflict Resolution}
\label{subsubsec:quantitative-denoising}

After hybrid recall, the candidate set faces two issues: intra-source noise, where semantic relevance scores follow a long-tail distribution with weakly relevant memories scattered in the low-score region, and cross-source conflicts, where different sources may describe the same event or stale abstractions contradict newer evidence. Mandol addresses both in two stages.

\paragraph{Intra-source denoising.}
Traditional static thresholds or methods based on mean and standard deviation assume a specific score distribution and are easily disturbed by extremely high or low scores, causing unstable filtering. Since different queries can produce non-normal score distributions, we apply a dynamic threshold based on the median absolute deviation (MAD), which robustly filters low-scoring noisy candidates without assuming normality:
\begin{equation}
    \tau_{mad} = \mathrm{median}(S_{ce}) - \kappa \cdot \mathrm{MAD}(S_{ce}),
    \label{eq:mad-threshold}
\end{equation}
where \(\mathrm{MAD}(S_{ce}) = \mathrm{median}(|S_{ce}(c_i) - \mathrm{median}(S_{ce})|)\). Candidates below \(\tau_{mad}\) are discarded. For example, a query to the episodic memory recalls 50 candidates, with a median \(S_{ce}\) score of 0.8 and a computed MAD of 0.1. With \(\kappa=2.5\), the dynamic threshold becomes \(0.8 - 0.25 = 0.55\), and only candidates with scores above 0.55 proceed to the next stage.


\paragraph{Cross-source conflict resolution.}
 After intra-source denoising, candidates from different sources may still conflict or contain redundancy. The system compares memory unit content across sources: if units involve the same entities but give inconsistent descriptions, or if abstract nodes trace back to the same basic memory identifiers, they are marked as conflicting. For each conflicting candidate \(c_i\), Mandol computes an arbitration score \(S_{arb}(c_i)\) that balances semantic relevance \(S_{ce}(c_i)\), temporal freshness \(f_{time}(c_i)\), and source confidence \(f_{source}(c_i)\) of the candidate $c_i$:

\begin{equation}
S_{arb}(c_i) = w_{rel} \cdot S_{ce}(c_i) + w_{temp} \cdot f_{time}(c_i) + w_{source} \cdot f_{source}(c_i),
\label{eq:arb}
\end{equation}
where \(f_{time}(c_i)\) models an exponential decay based on the elapsed time, giving higher scores to more recent information. \(f_{source}(c_i)\) assigns a confidence weight according to the memory source. Specifically, basic memory receives the highest weight (1.0) because it contains raw dialogue information. Episodic memory receives a medium weight (0.8), and semantic memory receives the lowest weight (0.6), as they have different levels of abstraction.
Mandol greedily keeps the highest-scoring record and discards stale or redundant alternatives. 
Continuing with the example of Mike's trip to Beijing, an emotional memory records an older preference ``\textit{likes coffee}'', while a recent basic memory unit captures ``\textit{I tried Longjing tea at a teahouse and loved it. I want to explore more Chinese teas}''. For a drink recommendation query, the basic memory observation is preferred because of its higher freshness and source confidence, overriding the outdated coffee preference.

\subsubsection{Token-Constrained High-Quality Context Generation}
\label{subsubsec:token-constrained-context}

After conflict resolution, the remaining candidate set may still exceed the token budget. Simple top-\(K\) truncation may retain candidates with low information density or drop critical supplementary evidence. To construct a compact yet comprehensive context, Mandol jointly optimizes relevance and diversity under the token budget, prioritizing highly relevant memories while preserving complementary evidence from different sources.

We adopt a greedy iterative selection algorithm based on joint optimization of relevance and diversity, introducing a maximal marginal relevance (MMR) objective:
\begin{equation}
    \operatorname{MMR}(c_i \mid S) = \lambda \cdot \operatorname{Rel}(c_i) - (1-\lambda) \cdot \operatorname{Red}(c_i, S),
    \label{eq:mmr}
\end{equation}
where \(\operatorname{Rel}(c_i)\) is the cross-encoder relevance score \(S_{ce}(c_i)\), and \(\operatorname{Red}(c_i, S)\) measures the maximum redundancy between the candidate \(c_i\) and the already selected set \(S\), based on entity and memory source overlap. The algorithm greedily adds candidates with the highest MMR score until the token budget is reached. This favors complementary evidence: if ``\textit{Beijing trip booking successful}'' is already selected, a near-duplicate ``\textit{Beijing trip order confirmed}'' receives a high redundancy penalty, while ``\textit{flight delayed due to weather}'' adds new information and is preferred.

\section{Implementation}
\label{sec:implementation}

\begin{table}[t]
\centering
\caption{Core APIs exposed by Mandol.}
\label{tab:api}
\begingroup
\footnotesize
\setlength{\tabcolsep}{3pt}
\renewcommand{\arraystretch}{1.08}
\begin{tabularx}{\columnwidth}{@{}>{\raggedright\arraybackslash}p{0.34\columnwidth}>{\raggedright\arraybackslash}X@{}}
\toprule
\textbf{Scope} & \textbf{API} \\
\midrule
Memory unit operation &
\texttt{add/delete/update(space, [unit])} \\

Explicit relationship &
\texttt{add/delete/update\_relationship(}\newline
\texttt{unit\_src, unit\_target, [type])} \\

Unit retrieval & \texttt{search\_unit(query, memory\_space, type)} \\

Graph traversal &
\texttt{traverse\_explicit\_nodes(unit, [type])} \\

Semantic traversal &
\texttt{traverse\_implicit\_nodes(unit, [top\_k])} \\

Quantitative retrieval &
\texttt{quantitative\_search(query, [params])} \\

Persistence &
\texttt{save\_graph([dir, build\_index])} \\

Memory construction &
\texttt{build\_memory\_from\_raw(sample\_id,}\newline
\texttt{extraction\_style, [session\_date])} \\
\bottomrule
\end{tabularx}
\endgroup
\end{table}

Mandol is implemented in Python as a single in-memory process with three modules: \emph{memory construction and storage}, \emph{memory retrieval}, and \emph{memory persistence}.
These modules build basic and high-level memories from raw dialogues, perform quantitative search for user queries, and persist/reload memories when needed.
Table~\ref{tab:api} summarizes the exposed retrieval, traversal, construction, and persistence APIs.

\paragraph{Memory construction and storage.}
When new dialogue data arrives, Mandol first partitions the raw dialogue into dialogue chunks, where each chunk corresponds to a local segment of a session.
For each chunk, Mandol extracts fine-grained memory units, including dialogue content, speaker information, and temporal information.
Mandol further builds contextual edges and reference relations among these units across sessions, forming a structured semantic graph in the basic memory space.
Based on this basic graph, Mandol constructs high-level memories in different memory spaces.
For episodic memory, Mandol extracts events, user states, and temporal relationships. Mandol then organizes them into temporal event chains stored in the episodic memory space.
For semantic memory, Mandol first extracts entities and relations, and then represents them as a structured semantic graph stored in the entity-relation memory space.
For emotional memory, Mandol extracts user preferences, long-term states, and preference changes, and organizes them into preference-evolution chains.
All high-level memory nodes maintain source links to their supporting basic memory units, allowing abstract memories to be traced back to the original dialogue evidence.
When a new dialogue arrives, Mandol incrementally updates the affected basic units, semantic graph edges, temporal chains, entity relations, and emotional memories rather than reconstructing the entire memory space.

\paragraph{Memory retrieval.}
Given a user query, Mandol invokes the quantitative search API to execute the quantitative query mechanism described in Section~\ref{subsec:smart-quantitative-querying} and produce the final memory context for answer generation.
It first analyzes the query intent and routes the query to suitable memory spaces, such as basic, episodic, emotional, or entity-relation memories.
Within each selected memory space, Mandol invokes \texttt{search\_unit} to retrieve relevant memory units.
It can then call the explicit-node or implicit-node traversal APIs to expand the candidates through graph edges or semantic neighbors.
This graph-augmented expansion improves evidence completeness for multi-hop and context-dependent queries.
Mandol then fuses candidates from different memory sources, removes noisy records, resolves stale or conflicting evidence, and selects high-quality memories under the token budget.
The final output is a compact LLM context containing relevant memory content.

\paragraph{Memory persistence.}
By default, Mandol keeps basic and abstract memories in the in-memory SemanticMap/Graph to support low-latency retrieval.
Mandol can also persist selected memory units and graphs for checkpointing, recovery, and later reloading into the in-memory runtime.
When memory usage exceeds a predefined threshold, Mandol selects less active memory units and relationship records using LRU/LFU-based policies, migrates their contents to the database via the \texttt{save\_graph} API, and keeps lightweight references in the memory graph.
If a later retrieval accesses a referenced but unloaded memory record, Mandol invokes a page-in callback to load the corresponding record from the database and rebuild the related units and edges in the in-memory graph.
Our current implementation uses DuckDB~\cite{raasveldt2019duckdb} as an embedded in-process persistence backend to store heterogeneous memory information, including vectors and graph relationships. 
In future work, we will also extend Mandol to other vector or graph databases.

\section{Evaluation}
\label{sec:evaluation}

In this section, we evaluate Mandol along three dimensions: memory retrieval quality, system performance, and resource utilization.
First, we compare Mandol with existing memory systems in terms of retrieval accuracy and token consumption.
Second, we compare system performance by measuring the latency of Search and Add operations, covering both memory retrieval and memory insertion.
Third, we study Mandol's resource utilization, including system RAM and GPU memory consumption.
These three aspects are reported in Sections~\ref{subsec:accuracy-evaluation}, \ref{subsec:latency-throughput-evaluation}, and~\ref{subsec:resource-deployment-evaluation}, respectively.

\begin{table*}[!t]
\centering
\caption{LoCoMo accuracy (\%) comparison among different memory systems. $\dagger$ denotes our reproduced results.}
\label{tab:locomo_main}
\begingroup
\footnotesize
\setlength{\tabcolsep}{0pt}
\renewcommand{\arraystretch}{1.02}
\begin{tabular*}{0.88\textwidth}{@{\extracolsep{\fill}}llcccccc@{}}
\toprule
\textbf{Backbone} & \textbf{System} & \textbf{Avg.\ Tok.} & \textbf{Single} & \textbf{Multi} & \textbf{Temp.} & \textbf{Open} & \textbf{Overall} \\
\midrule
\multirow{6}{*}{\textit{GPT-4o-mini}}
& Mem0 & 1.0k & 66.71 & 58.16 & 55.45 & 40.62 & 61.00 \\
& MemU & 4.0k & 72.77 & 62.41 & 33.96 & 46.88 & 61.15 \\
& MemOS & 2.5k & 81.45 & 69.15 & 72.27 & 60.42 & 75.87 \\
& Zep & 1.4k & 88.11 & 71.99 & 74.45 & 66.67 & 81.06 \\
& EverMemOS$^\dagger$ & 2.5k & \underline{91.68} & \underline{82.74} & \underline{79.34} & \textbf{70.14} & \underline{86.13} \\
& \textbf{Mandol (Ours)} & 2.0k & \textbf{93.82} & \textbf{85.11} & \textbf{89.10} & 65.63 & \textbf{89.48} \\
\midrule
\multirow{6}{*}{\textit{GPT-4.1-mini}}
& Mem0 & 1.0k & 68.97 & 61.70 & 58.26 & 50.00 & 64.20 \\
& MemU & 4.0k & 74.91 & 72.34 & 43.61 & 54.17 & 66.67 \\
& MemOS & 2.5k & 85.37 & 79.43 & 75.08 & 64.58 & 80.76 \\
& Zep & 1.4k & 90.84 & 81.91 & 77.26 & 75.00 & 85.22 \\
& EverMemOS$^\dagger$ & 2.3k & \underline{95.32} & \underline{89.01} & \textbf{90.13} & \underline{77.43} & \underline{91.97} \\
& \textbf{Mandol (Ours)} & 1.9k & \textbf{95.36} & \textbf{92.20} & 87.85 & \textbf{79.17} & \textbf{92.21} \\
\bottomrule
\end{tabular*}
\endgroup
\end{table*}

\begin{table*}[!t]
\centering
\caption{LongMemEval accuracy (\%) comparison among different memory systems.}
\label{tab:longmemeval_main}
\begingroup
\footnotesize
\setlength{\tabcolsep}{0pt}
\renewcommand{\arraystretch}{1.02}
\begin{tabular*}{0.98\textwidth}{@{\extracolsep{\fill}}llcccccccc@{}}
\toprule
\textbf{Backbone} & \textbf{System} & \textbf{Avg.\ Tok.} & \textbf{SS-Pref} & \textbf{SS-Asst} & \textbf{Temporal} & \textbf{Multi-S} & \textbf{Know.\ Upd.} & \textbf{SS-User} & \textbf{Overall} \\
\midrule
\multirow{5}{*}{\textit{GPT-4o-mini}}
& MemU & 0.5k & 76.70 & 19.60 & 17.30 & 42.10 & 41.00 & 67.10 & 38.40 \\
& Mem0 & 1.1k & \underline{90.00} & 26.78 & 72.18 & 63.15 & 66.67 & 82.86 & 66.40 \\
& Zep & 1.6k & 53.30 & \underline{75.00} & 54.10 & 47.40 & \underline{74.40} & 92.90 & 63.80 \\
& MemOS & 1.4k & \textbf{96.67} & 67.86 & \underline{77.44} & \underline{70.67} & 74.26 & \underline{95.71} & \underline{77.80} \\
& \textbf{Mandol (Ours)} & 2.1k & \textbf{96.67} & \textbf{98.21} & \textbf{78.95} & \textbf{74.44} & \textbf{88.46} & \textbf{97.14} & \textbf{85.00} \\
\midrule
\multirow{2}{*}{\textit{GPT-4.1-mini}}
& EverMemOS & 2.8k & \underline{93.33} & \underline{85.71} & \underline{77.44} & \underline{73.68} & \textbf{89.74} & \underline{97.14} & \underline{83.00} \\
& \textbf{Mandol (Ours)} & 2.3k & \textbf{96.67} & \textbf{98.21} & \textbf{87.22} & \textbf{77.44} & \textbf{89.74} & \textbf{98.57} & \textbf{88.40} \\
\bottomrule
\end{tabular*}
\endgroup
\end{table*}

\subsection{Experimental Setup}
\label{subsec:eval-setup}

\paragraph{Datasets and baselines.}
We evaluate Mandol on two widely used long-term conversational memory benchmarks.
\textbf{LoCoMo}~\cite{maharana2024evaluating} provides 10 long multi-session conversations with 1,986 questions and ground-truth answers, designed to test retrieval and reasoning over long conversational histories.
Following prior work, we evaluate four question types: single-hop recall, multi-hop reasoning, temporal reasoning, and open-domain grounding, corresponding to the Single, Multi, Temp., and Open columns in Table~\ref{tab:locomo_main}.
\textbf{LongMemEval}~\cite{wang2024augmenting} contains 500 manually curated questions over long user-assistant interaction histories.
Following prior work, we evaluate six question types: single-session user preferences, single-session assistant information, temporal reasoning, multi-session reasoning, knowledge update, and single-session user information, corresponding to the SS-Pref, SS-Asst, Temporal, Multi-S, Know. Upd., and SS-User columns in Table~\ref{tab:longmemeval_main}.
We compare Mandol with representative open-source memory systems, including Mem0~\cite{chhikara2025mem0}, MemU~\cite{nevamind2026memu}, Zep~\cite{rasmussen2025zep}, MemOS~\cite{memos2025}, and EverMemOS~\cite{evermemos2026}.

\paragraph{Metrics and environments.}
To evaluate retrieval quality, we use QA accuracy on the two benchmarks.
QA accuracy is defined as the percentage of questions whose generated answers are judged correct or semantically consistent with the ground-truth answers.
Following prior memory system evaluations, we use GPT-4o-mini and GPT-4.1-mini as the answer-generation backbones and adopt the LLM-based answer correctness evaluation script from EverMemOS~\cite{evermemos2026}.
To evaluate system performance, we follow the performance measurement setup of MemOS~\cite{memos2025} and measure API-level Search latency and Add latency.
Search latency measures the end-to-end time for retrieving relevant memories given an input query, while Add latency measures the online insertion time, including representation generation and index updates.
Latency is measured in milliseconds.
We conduct our main experiments on a cloud server equipped with one NVIDIA H800 GPU with 80~GB memory, 20 Intel Xeon Platinum 8458P vCPUs, and 120~GB system memory.
For additional local deployment analysis, we evaluate Mandol on a laptop equipped with an Intel Core Ultra 9 275HX CPU, 96~GB system memory, and an NVIDIA RTX 5090 Laptop GPU with 24~GB memory.

\subsection{Accuracy and Token Efficiency Results}
\label{subsec:accuracy-evaluation}

\begin{table*}[!t]
\centering
\caption{Latency comparison for memory retrieval (Search) and memory insertion (Add), measured in milliseconds.}
\label{tab:latency_main}
\begingroup
\footnotesize
\setlength{\tabcolsep}{4pt}
\renewcommand{\arraystretch}{1.06}

\begin{tabular*}{0.92\textwidth}{@{\extracolsep{\fill}}llcccccc@{}}
\toprule
\multirow{2}{*}{\textbf{Deployment}} &
\multirow{2}{*}{\textbf{System}} &
\multicolumn{3}{c}{\textbf{Search (Retrieval, ms)}} &
\multicolumn{3}{c}{\textbf{Add (Memory Insertion, ms)}} \\
\cmidrule(lr){3-5} \cmidrule(lr){6-8}
& &
\textbf{P99} &
\textbf{P90} &
\textbf{Mean} &
\textbf{P99} &
\textbf{P90} &
\textbf{Mean} \\
\midrule
\multirow{6}{*}{\textit{Server}}
& MemU                & 63000.7 & 60539.5 & 47554.5 & 12070.6 & 7273.1 & 5077.9 \\
& EverMemOS$^\dagger$ & 37192.4 & 35220.4 & 20092.1 & 790.2 & 555.5 & 317.7 \\
& Mem0                & 4637.0  & 1397.0  & 1089.0  & 2841.0 & 1650.0 & 888.0 \\
& Zep                 & 5348.7  & 614.8   & 571.7   & 375.1  & 254.5  & 239.0 \\
& MemOS               & 777.1   & 528.4   & 440.5   & 376.4  & 211.6  & 191.9 \\
& Mandol (Ours)       & \textbf{94.8} & \textbf{88.5} & \textbf{82.2} & \textbf{67.3} & \textbf{46.9} & \textbf{39.7} \\
\midrule
\textit{Local}
& Mandol (Ours)       & \textbf{211.6} & \textbf{186.8} & \textbf{166.5} & \textbf{51.6} & \textbf{42.1} & \textbf{37.4} \\
\bottomrule
\end{tabular*}

\vspace{2pt}
\begin{minipage}{0.92\textwidth}
\footnotesize
\setlength{\parindent}{0pt}
Server: NVIDIA H800 80GB, 10 QPS for Search/Add.
Local: RTX 5090 Laptop 24GB, 5 QPS for Search and 10 QPS for Add.
$\dagger$ Reproduced using the official implementation.
\end{minipage}
\endgroup
\end{table*}

Tables~\ref{tab:locomo_main} and~\ref{tab:longmemeval_main} summarize the accuracy and average token usage on LoCoMo and LongMemEval.
In these tables, $\dagger$ marks our reproduced EverMemOS results; \textbf{bold} denotes the best result; underlined values denote the second-best result.
Mandol achieves the best overall accuracy on both benchmarks under all reported backbone settings. On LoCoMo, Mandol improves the strongest baseline by 3.35 percentage points under GPT-4o-mini and by 0.24 percentage points under GPT-4.1-mini, reaching 89.48\% and 92.21\% overall accuracy, respectively. On LongMemEval, Mandol improves the strongest baseline by 7.20 percentage points under GPT-4o-mini and by 5.40 percentage points under GPT-4.1-mini, reaching 85.00\% and 88.40\% overall accuracy, respectively.

Across question categories, Mandol is particularly effective for reasoning and temporal memory queries. On LoCoMo, Mandol achieves the best accuracy on single-hop and multi-hop questions under both backbones, and also obtains the best temporal-reasoning accuracy under GPT-4o-mini. These results show that Mandol can retrieve direct facts while composing evidence across long interaction histories. On LongMemEval, Mandol obtains notable gains on single-session information tracking, temporal reasoning, and multi-session reasoning. These tasks require the system to track information provided by the assistant, recognize temporal cues, and combine evidence scattered across multiple sessions.
Mandol also improves knowledge-update accuracy, suggesting that its memory organization can better preserve evolving user states.

Mandol performs slightly worse than EverMemOS in only two cases on LoCoMo: open-domain questions under GPT-4o-mini and temporal reasoning under GPT-4.1-mini. This is largely because EverMemOS relies on larger retrieval models (Qwen3-Embedding-4B and Qwen3-Reranker-4B), whereas Mandol adopts a lightweight backend (Qwen3-Embedding-0.6B and bge-reranker-v2-m3~\cite{chen2024bgem3}). Despite this lighter backend, Mandol still achieves higher overall accuracy and consistently outperforms EverMemOS across all LongMemEval cases. Mandol also consumes fewer tokens, reducing token usage by 17.4\%--20.0\% compared to EverMemOS. This accuracy–efficiency pattern suggests that the gains mainly come from Mandol's memory organization and quantitative retrieval mechanism rather than from the retrieval backend models.

We attribute this overall advantage to two design choices. First, Mandol uses a unified structured memory representation to organize basic memory units, episodic events, entity relations, temporal links, user states, and high-level memories. This design improves evidence completeness and traceability, especially for queries that require evidence from multiple related records across sessions. Second, Mandol uses a quantitative query mechanism to select suitable memory sources, filter noisy candidates, resolve conflicting evidence, and construct a higher-quality context under a limited token budget. These two designs make Mandol particularly effective for multi-hop, temporal, and update-sensitive queries.

\begin{figure}[t]
    \centering
    \includegraphics[width=0.95\columnwidth]{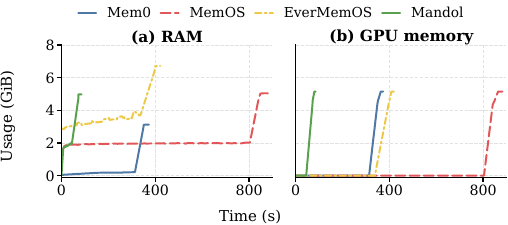}
    \Description{Resource-usage plot comparing RAM and GPU memory usage across memory systems during the end-to-end workload.}
    \caption{Resource usage during the end-to-end workload.}
    \label{fig:resource_usage}
\end{figure}

\subsection{System Performance Results}
\label{subsec:latency-throughput-evaluation}

For latency evaluation, we measure API-level Search and Add operations on LoCoMo. Search latency is the time from the arrival of a query to the return of the retrieved memory context, excluding downstream answer generation. Add latency measures the time for inserting a new memory item, including representation generation, index updates, and graph updates. QPS denotes the request rate used in the serving workload. For example, 10 QPS means that the system receives 10 input queries per second. For each query, we report the P99, P90, and mean latency. P99 and P90 denote the 99th- and 90th-percentile latency, which characterize tail latency under concurrent requests. For comparability, we report single-pass retrieval latency for all systems. In particular, EverMemOS requires multi-round retrieval to reach its reported accuracy, but here we only measure the latency of one retrieval pass, aligning with the other systems.

Table~\ref{tab:latency_main} summarizes the latency results in milliseconds. On the server setting, Mandol achieves the best Search latency across P99, P90, and mean latency. Compared with the second-fastest system MemOS, Mandol reduces P99 Search latency from 777.1~ms to 94.8~ms, achieving about an 8.2$\times$ speedup, and reduces mean Search latency from 440.5~ms to 82.2~ms, achieving about a 5.4$\times$ speedup. Mandol also achieves the best Add latency. For P99 Add latency, compared with the second-fastest system, Zep, Mandol achieves about a 5.6$\times$ speedup. For mean Add latency, compared with the second-fastest system MemOS, Mandol achieves about a 4.8$\times$ speedup. 
The local deployment results further show that Mandol remains efficient even on a laptop. Its local Search latency and Add latency are even lower than the server-side latency of MemOS.

The latency advantage mainly comes from three factors. First, many existing memory systems rely on external database services, so each query requires serialization, database execution, result transfer, and parsing before the Python-side pipeline can continue. This database round trip becomes a substantial overhead on the serving path. Second, multi-store systems further introduce cross-database orchestration overhead, such as retrieving candidates from a vector store and then issuing additional filtering, expansion, or relation queries to a graph or relational store. These steps increase both tail latency and mean latency. Third, Mandol adopts an in-process memory architecture, where retrieval indexes, structured memory links, and active memory states are maintained within the same runtime process. This design enables direct memory access and lightweight function calls for retrieval and update operations, avoiding most database round trips and reducing the cost of both retrieval and online memory insertion.

\subsection{Resource Consumption}
\label{subsec:resource-deployment-evaluation}

In addition to latency, we measure RAM and GPU memory consumption over the end-to-end LoCoMo workload, including memory insertion, index construction/update, and a small retrieval workload. This captures the runtime footprint of each deployed system, including resident database services, in-process caches, acceleration libraries, and active retrieval components.

Figure~\ref{fig:resource_usage} reports processing time and memory consumption under the same lightweight retrieval backend: Qwen3-Embedding-0.6B for embedding and bge-reranker-v2-m3 for reranking.
RAM usage varies more substantially across systems. Mem0 has the lowest RAM usage because it uses a lightweight vector-database stack. MemOS maintains both vector and graph database components, while EverMemOS has the largest footprint due to multiple active database components and indexes. Mandol has a moderate footprint because it keeps memory data, retrieval indexes, cached representations, and structured links inside the native in-memory process. 
GPU memory usage is similar across systems because the loaded embedding and reranking models are the same under the unified retrieval-model setting. Finally, Mandol completes the LoCoMo workload in 86.18~s, 4.2--9.9$\times$ faster than the compared systems. This speedup mainly comes from its in-memory construction and update pipeline, which avoids external database writes and cross-store coordination.

\section{Conclusion}
\label{sec:conclusion}

This paper presents Mandol, an agglomerative memory system for long-term conversations, which consolidates fragmented memory representations and storage into a unified, memory-native architecture.
Mandol combines a hierarchical memory model, agglomerative semantic data structure, and quantitative retrieval to unify memory representation, reduce storage fragmentation, support low-latency hybrid queries, and construct accurate token-efficient contexts.
Experiments on long-conversation benchmarks show that Mandol improves retrieval quality and system performance over existing agent memory systems while maintaining low latency on consumer-grade hardware.

\bibliographystyle{ACM-Reference-Format}
\bibliography{references}

\end{document}